# Mechanical Properties of Graphene Nanowiggles

R. A. Bizao, T. Botari and D. S. Galvao
Applied Physics Department, State University of Campinas, 13083-970, Campinas-SP, Brazil

**ABSTRACT**

In this work we have investigated the mechanical properties and fracture patterns of some graphene nanowiggles (GNWs). Graphene nanoribbons are finite graphene segments with a large aspect ratio, while GNWs are nonaligned periodic repetitions of graphene nanoribbons. We have carried out fully atomistic molecular dynamics simulations using a reactive force field (ReaxFF), as implemented in the LAMPPS (Large-scale Atomic/Molecular Massively Parallel Simulator) code. Our results showed that the GNW fracture patterns are strongly dependent on the nanoribbon topology and present an interesting behavior, since some narrow sheets have larger ultimate failure strain values. This can be explained by the fact that narrow nanoribbons have more angular freedom when compared to wider ones, which can create a more efficient way to accumulate and to dissipate strain/stress. We have also observed the formation of linear atomic chains (LACs) and some structural defect reconstructions during the material rupture. The reported graphene failure patterns, where zigzag/armchair edge terminated graphene structures are fractured along armchair/zigzag lines, were not observed in the GNW analyzed cases.

**INTRODUCTION**

The advent of graphene [1] opened a new era in materials science. Graphene has unique electronic, optical and mechanical properties [2]. Because of these exceptional properties, graphene is considered as the basis for a new nanoelectronics. However, graphene is a zero bandgap semiconductor in its pristine form, which poses limitations to its use in some kind of transistors. One possible way to open the graphene gap is to create graphene nanoribbons [3,4]. Graphene nanoribbons are finite graphene segments with a large aspect ratio.

Recently, a new family of nanoribbons, called graphene nanowiggles (GNWs) [5] has been object of many experimental and theoretical investigations [6,7]. GNWs are nonaligned periodic repetitions of graphene nanoribbons as can be seen on Figure 1(a). In Figure 1(b-d) GNW structures are shown in yellow and the 'cut' region to form a GNW from a graphene sheet is shown in red.

GNWs were synthetized by a recently bottom-up technique, which allows a precise control of the generated structure, instead of the more standard synthesis techniques, such as, chemical or lithographic methods [5], where fine control is difficult to achieve.

One of the ways to uniquely characterize the GNW geometry is through the definition of four structural parameters: $P_\alpha$, $O_\beta$, $L_P$ and $L_o$. The α and β indices can have **A** or **Z** values, depending on whether the segment if armchair (**A**) or a zigzag (**Z**) one. The $P_\alpha$ and $O_\beta$ parameters characterize the GNW segment width by the number of C-C dimer lines (α or β = **A**), or by zigzag stripes (α or β = **Z**), for parallel ($P_\alpha$) and oblique ($O_\beta$) segments, respectively [7].

In this sense, there are four possible combinations of $P_\alpha$ and $O_\beta$, which can originate four possible GNW families: **AA** (Figure 1(a)), **ZA** (Figure 1(b)), **AZ** (Figure 1(c)), and **ZZ** (Figure 1(d)). The $L_P$ parameter characterizes the inner parallel segment and $L_o$ the width of the original GNW, that is to say, before cutting the trapezoids to form a GNW from its parent graphene sheet.

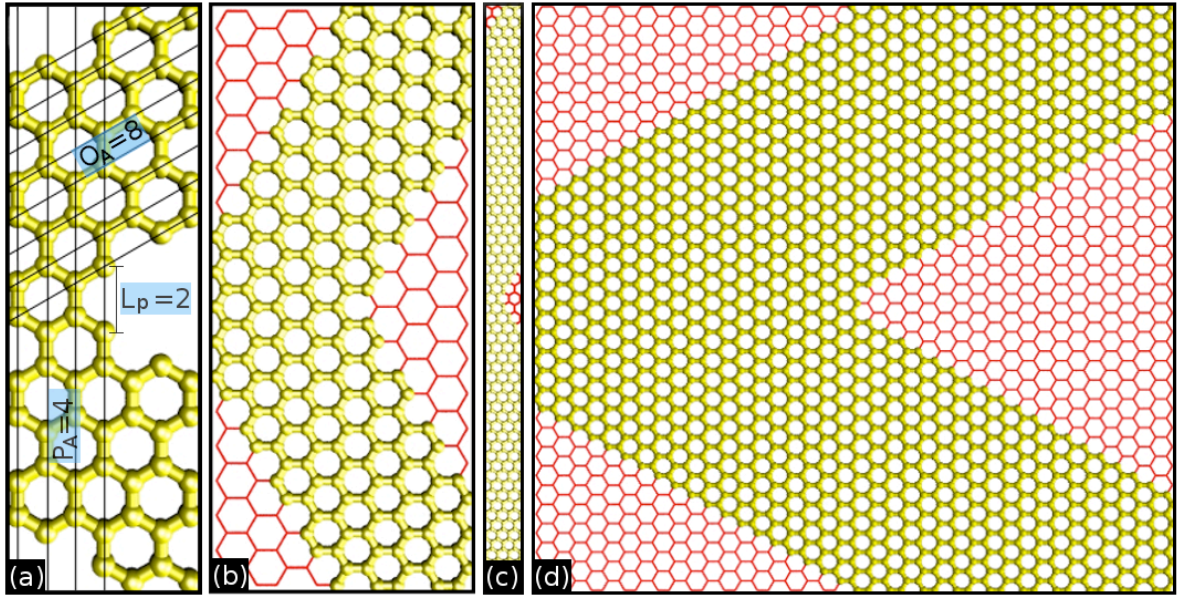

**Figure 1.** Examples of unit cells for $(P_\alpha, O_\beta)$ = (a) $(4_A, 8_A)$; (b) $(5_Z, 10_A)$; (c) $(4_A, 15_Z)$; and (d) $(20_Z, 15_Z)$ GNWs. In Figure 1(a) it is shown a pristine GNW, while in Figure 1(b-d) the 'cut' regions are shown in red to form a GNW from its parent graphene sheet. See text for discussions.

It was recently reported that GNW electronic and magnetic properties can be tunable by the geometrical variation of its structure and that this could be exploited to create new devices [6,7]. Although their electronic and magnetic properties have already been studied in details, the study of their mechanical properties and fracture patterns under mechanical load is still missing. This kind of study is one of the objectives of this present work.

In this work we have investigated the mechanical properties and fracture patterns of a family of GNWs of different topologies through fully atomistic molecular dynamics simulations.

**METHODOLOGY**

We have carried out molecular dynamics simulations by using the reactive force field ReaxFF [8], as implemented in the LAMMPS (Large-scale Atomic/Molecular Massively Parallel Simulator) package [9]. The simulations were performed at 150 K with a Nosé-Hoover thermostat and an accurate time step of 0.05 fs.

ReaxFF is a reactive force field that can simulate the formation and dissociation of bonds between atoms. This force field parameter calibration is made using very accurate DFT calculations. The obtained ReaxFF data are then contrasted against experimental data and the process iterated until convergence [8]. ReaxFF has been successfully used to investigate large reactive structures when *ab initio* methods become computational prohibitive [10,11].

The GNW structures we investigated here contain 2000 atoms on average; the structures were generated by unit cell replication. Due to space limitations, the discussion was restricted to the simulations carried for AZ GNWs. Similar results were obtained to other families. The parameter $L_p$ was fixed to provide the smallest possible inner parallel segment, that is $L_p = 2$

(for AZ GNW), and the parameter $L_o$ was fixed to $L_o=2P_\alpha - 1$ [7]. Other geometrical specifications to obtain the needed GNW unit cells are described in [7].

After the structure generation, we then proceed to carry out the MD simulations. Firstly, we run some simulation steps at NPT ensemble in order to provide a thermal stress relaxation. Then, we run MD simulations at NVT ensemble and stress/strain calculation were performed.

In order to obtain the stress strain curves, we stretched the structures with a strain rate of $10^{-5}$ fs along its periodic direction and then the stress response is calculated. The strain was imposed by a gradual increase of the super cell dimensions and the stress response was obtained from the calculated forces for each atom divided by the total volume of the structure considering the well-established graphene van de Walls thickness of 3.4 Å.

We have also calculated the von Mises stress for each atom to analyze the stress distribution on the structure under strain. The von Mises stress provides helpful information about the fracture process, since we can estimate the mechanical load through the structure [12].

**RESULTS AND DISCUSSIONS**

In order to obtain non-stretched structures, as mentioned in the methodology section, initially thermal relaxations were performed and during this process some bond reconstruction along the edges and also mechanical ripples along the periodic direction were observed. However, when the strain began to be applied the ripples disappear and the stress values begin to increase. For small strain values, the stress increases linearly and the Young's modulus can be calculated by a linear fitting. The linear behavior is characteristic of stress-strain elastic region and corresponds to a reversible process, that is to say, if the strain load is continuously decreased (and consequently the stress) until it reaches a zero value, the original structure is recovered [12]. If the stress continues to increase, there is a point at which the stress-strain curve is not linear anymore, characterizing the non-linear region, also known as plastic region. The strain in the plastic region is not reversible and when the strain load becomes zero, the structure remains deformed due to, for example, some bond reconstructions in the structure. Continuing the strain increase, the fracture process begins and the stress decreases quickly.

The stress-strain curves for: (a) ($4_A,10_Z$); (b) ($6_A,14_Z$), and; ($13_A,22_Z$), all of them from the AZ GNW family, are shown in Figure 2. The estimated Young's modulus (Y) values obtained from the linear fitting in the elastic region resulted in: (a) Y = 759(4), (b) Y = 797(7) and (c) Y = 577(5) GPa.

The stress-strain processes during the fractures showed that the strain goes down quickly and some oscillations of the stress values are observed after the fractures. These fluctuations occur due to energy accumulation on the structure during the strain application and when the structure is fractured the accumulated energy is released and the oscillations can happen. The fracture processes occur approximately with 0.1 of strain load values for the analyzed structures. We can define the ultimate strength (US) point, characterized by a critical stress ($\sigma_c$) and a corresponding critical strain ($\varepsilon_c$), as the highest strength that the material can support before its rupture. These US points are indicated in Figure 2 by the arrows. The obtained values of $\sigma_c$ were: $\sigma_c \cong$ (a) 72.12; (b) 62.54, and; (c) 53.94 GPa and the correspondents $\varepsilon_c$ were $\varepsilon_c \cong$ (a) 0.11, (b) 0.09 and (c) 0.10. Both Young's modulus and ultimate strength values are smaller when compared to graphene nanoribbon ones [13], showing that the crack plays an essential role to produce poorer mechanical response in structures like these.

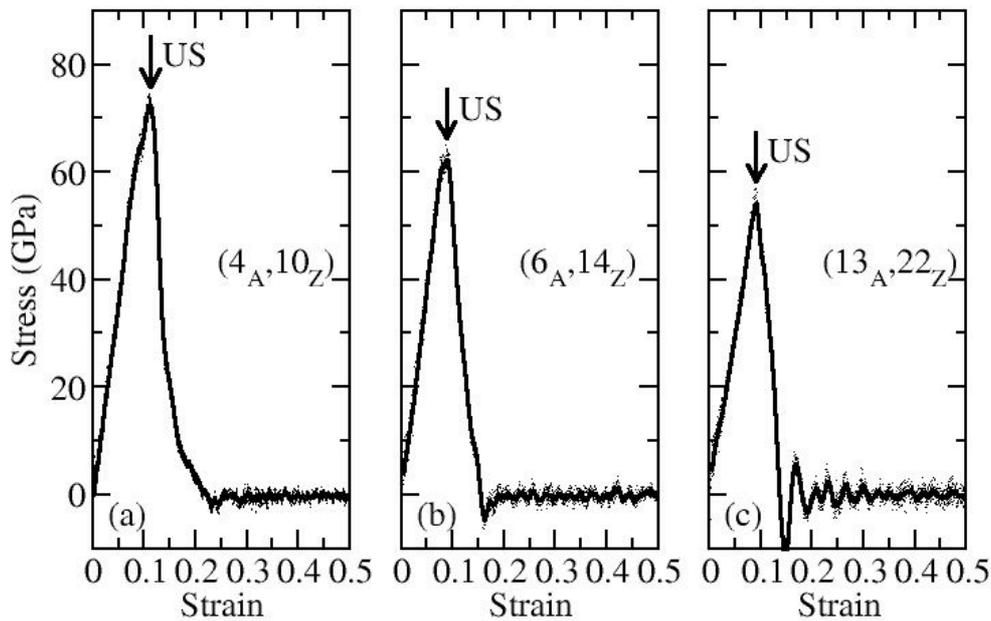

**Figure 2.** Stress-strain curves for: (a) $(4_A,10_Z)$; (b) $(6_A,14_Z)$, and; $(13_A,22_Z)$ AZ GNWs.

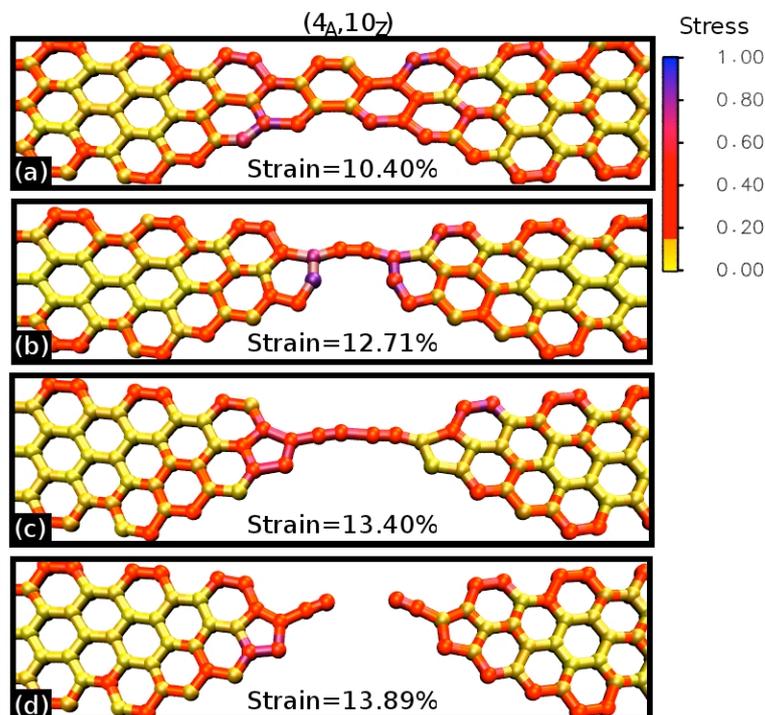

**Figure 3.** Fracture dynamics of the $(4_A,10_Z)$ GNW showing the stress distribution with a pallet color. In the beginning of the process the stress is concentrated at the borders of the structure. When the strain increases, a higher stress is located at the middle of the structure.

The von Mises stress distribution was calculated and used to investigate the fracture processes. We observed that first the stress accumulates on the borders of the structures, as shown in Figure 3 (a). As times goes on, the stress begins to accumulate in a reduced part of the structure until fracture happens, characterized by breaking bonds in this region. During this process some reconstructions can be observed, as well as, the formation of pentagons. Also, the formation of linear atomic chains (LACs) were observed during the fracture, as can be seen during the time evolution simulation snapshots of a ($4_A$,$10_Z$) GNW under strain, illustrated in Figure 3.

## CONCLUSIONS

We have investigated through reactive molecular dynamics simulations the mechanical properties of the graphene nanowiggles under mechanical stress/strain. We observed during the relaxation process some edge reconstruction, as well as, structural ripples along the periodic direction. We calculate the Young's modulus by a linear fitting of the stress strain curves at the elastic region and characterized the stress process. Also, during the stress strain simulations we observed during the fracture the creation of some linear atomic chains formations and bond reconstructions along the GNW membrane.

Our results show that the GNW fracture patterns present an interesting behavior, in the sense that some narrow GNWs have higher ultimate failure strength values than the wider ones. This can be explained by the fact that narrow nanoribbons have more angular freedom when compared to wider ones, which can create a more efficient way to accumulate and to dissipate strain/stress. Also, the reported graphene failure patterns, where zigzag/armchair edge terminated graphene structures are fractured along armchair/zigzag lines, were not observed in the analyzed GNW cases.

## ACKNOWLEDGMENTS

This work was supported in part by the Brazilian Agencies CNPq, CAPES and FAPESP. The authors would like to thank the Center for Computational Engineering and Sciences at Unicamp for financial support through the FAPESP/CEPID Grant 2013/08293-7.

## REFERENCES

1. K. S. Novoselov *et al.*, *Science* **306**, 666 (2004).
2. A. K. Geim and K. S. Novoselov, *Nature Materials* **6**,183 (2007).
3. K. Kim *et al.*, *Nano Lett.* **12**, 293 (2011).
4. I. Deretzis and A. La Magna, *Eur. Phys. J. B* **81**, 15 (2011).
5. J. Cai *et al.*, *Nature* **466**, 470 (2010).
6. E. C. Girao *et al.*, *Phys. Rev. Lett.* **107**, 135501 (2011).
7. E. C. Girao *et al.*, *Phy. Rev. B* **85**, 235431 (2012).
8. A. C. T. van Duin *et al.*, *J. Phys. Chem. A* **105**, 9396 (2001).
9. S. Plimpton, *J. Comp. Phys.* **117**, 1 (1995).
10. M. Z. S. Flores, P. A. S. Autreto, S. B. Legoas, and D. S. Galvao, *Nanotechnology* **20**, 465704 (2009).
11. R. Paupitz *et al.*, *Nanotechnology* **24**, 035706 (2012).
12. M. J. Buehler, *Atomistic Modeling of Materials Failure*, Springer, New York (2008).
13. R. Faccio *et al., J. Phys.: Condens. Matter* **21**, 285304 (2009).